%% file: haze7.tex
\documentclass[numberedappendix]{emulateapj}
\usepackage{graphicx}
\newcommand{\degree}{^\circ}

\newcommand{\be}{\begin{equation}}
\newcommand{\ee}{\end{equation}}
\newcommand{\bp}{\begin{figure}[!ht]}
\newcommand{\ep}{\end{figure}}
\newcommand{\bpm}{\begin{figure*}[!ht]}
\newcommand{\epm}{\end{figure*}}
\newcommand{\reffig}[1]{Figure \ref{fig:#1}}
\newcommand{\refsec}[1]{\S \ref{sec:#1}}

\begin{document}


\title{A Last Look at the Microwave Haze/Bubbles with WMAP}

\author{Gregory Dobler\altaffilmark{1,2}}

\altaffiltext{1}{
  Kavli Institute for Theoretical Physics,
  University of California, Santa Barbara
  Kohn Hall, Santa Barbara, CA 93106 USA
}
\altaffiltext{2}{
  dobler@kitp.ucsb.edu
}


\begin{abstract}
  \input{abstract}
\end{abstract}


\section{Introduction}
\label{sec:introduction}
  \input{introduction}

\section{Component Separation}
\label{sec:methods}
  \input{methods}

\section{Spectrum and Morphology}
\label{sec:specandmorph}
  \input{specandmorph}

\section{Polarization}
\label{sec:polarization}
  \input{polarization}

\section{Origin Scenarios}
\label{sec:interpretation}
  \input{interpretation}

\section{Summary}
\label{sec:summary}
  \input{summary}

\input{acknowledgements}


\input{haze7bib}
\end{document}

%% file: abstract.tex
The microwave ``haze'' was first discovered with the initial release
of the full sky data from the Wilkinson Microwave Anisotropy Probe.
It is diffuse emission towards the center of our Galaxy with spectral
behavior that makes it difficult to categorize as any of the
previously known emission mechanisms at those wavelengths.  With now
seven years of WMAP data publicly available, we have learned much
about the nature of the haze, and with the release of data from
the \emph{Fermi Gamma-Ray Space Telescope} and the discovery of the
gamma-ray haze/bubbles, we have had a spectacular confirmation of its
existence at other wavelengths.  As the WMAP mission winds down and
the \emph{Planck} mission prepares to release data, I take a last look
at what WMAP has to tell us about the origin of this unique Galactic
feature.  Much like the gamma-rays, the microwave haze/bubbles is
elongated in latitude with respect to longitude by a factor of roughly
two, and at high latitudes, the microwave emission cuts off sharply
above $\sim$35 degrees (compared to $\sim$50 degrees in the gammas).
The hard spectrum of electrons required to generate the microwave
synchrotron is consistent with that required to generate the gamma-ray
emission via inverse Compton scattering, though it is likely that
these signals result from distinct regions of the spectrum ($\sim$10
GeV for the microwaves, $\sim$1 TeV for the gammas).  While there is
no evidence for significant haze polarization in the 7-year WMAP data,
I demonstrate explicitly that it is unlikely such a signal would be
detectable above the noise.

%% file: introduction.tex
With the initial release of the full sky data by the
\emph{Wilkinson Microwave Anisotropy Probe} \citep[WMAP,][]{bennett03a} 
came the discovery that there was anomalously hard spectrum microwave
emission towards the center of our Galaxy \citep{finkbeiner04a}.  The
method for uncovering this WMAP ``haze'' as it was termed at the time
was very simple: there were known or anticipated microwave emission
mechanisms from the Galaxy and it was believed that templates (maps at
other wavelengths) morphologically tracing these foregrounds could be
used to clean Galactic emission from the maps for CMB analysis and to
study the foregrounds in their own right \citep{bennett03b} using
straightforward linear regression techniques.  These templates were
known to be insufficient down to the measurement noise in the WMAP
maps, yet the template fitting procedures worked remarkably well for
both the cosmological \citep{spergel03} and Galactic science purposes.

However, while removing $>$95\% of the variance in the data
\citep{finkbeiner04a,dobler08a} the templates failed most spectacularly 
within $\sim30\degree$ of the Galactic center (GC).  After removing
the emission correlated with the templates, the residual resembled a
``hazy'' blob with a spectrum that was too soft to be free-free
emission and too hard to be synchrotron emission from electrons
generated by supernova (SN) remnants \citep[][hereafter
DF08]{dobler08a}.  This microwave haze is somewhat elongated in
Galactic latitude with an intensity falling off with distance from the
GC.  A careful measurement of the spectrum and radial dependence of
the haze by DF08 led to ample speculation about the origin of the haze
electrons: from star formation and SN acceleration
\citep{biermann10} to signatures of particle dark matter
annihilation in our Galactic halo \citep{hooper07,cholis09a,cholis09b}
to claims that the feature did not exist at all \citep{mertsch10}.

The last of these claims was laid to rest with the first year data
release of the \emph{Fermi Gamma-Ray Space Telescope} which showed a
similar structure at gamma-ray energies \cite{dobler10}.  The
implication was that the same electrons that generate microwaves at
WMAP wavelengths generate gammas via inverse Compton (IC) scattering
that are observed by \emph{Fermi}.  It was also found
by \cite{dobler10} that, while the extent of the microwaves is
$\sim\pm30\degree$, the gamma-ray haze extended to a full
$\pm50\degree$ above and below the Galactic plane, making it the
second largest structure in our Galaxy after the Galactic disk
(assuming that it is centered roughly on the GC) with a top-to-bottom
length of 20 kpc.  Furthermore, analysis of 1.6 years of \emph{Fermi}
data by \cite{su10} suggests that the gamma-ray emission also has
sharp edges and it was renamed the ``bubbles''.  Given the established
correspondence between the WMAP ``haze'' and the \emph{Fermi}
``haze/bubbles'', I will refer to the microwave emission as the WMAP
``haze/bubbles'' or ``haze'' interchangeably throughout this
paper.\footnote{Historically, there has been some ambiguity to the
nomenclature and to what the term ``haze'' refers.  Given the
observations, the clearest definition would be that the haze is a
population of anomalously hard spectrum cosmic-rays towards the GC and
the microwave and gamma-ray haze/bubbles are observed signals
generated by those cosmic-rays.}

We now have the benefit of six additional years of microwave data from
WMAP and clues from the \emph{Fermi} gammas.  Given that the release
of the \emph{Planck} data is scheduled for late 2012, I will present a
description of what we know about the haze/bubbles at the end of the
WMAP era and the beginning of the \emph{Planck} era.  In particular,
in \refsec{methods} I will describe the methods used to uncover the
haze in the first place and comment on other component separation
techniques.  In \refsec{specandmorph} I present the 7-year WMAP
haze/bubbles spectrum and morphology and describe new findings at high
latitude.  In \refsec{polarization} I address the state of the haze in
the WMAP polarization data and in \refsec{interpretation} I discuss
the merits and demerits of the various proposals for the origin of the
haze/bubbles electrons before summarizing in \refsec{summary}.

%% file: methods.tex
Component separation in the context of CMB foregrounds is a broad term
for identifying and removing (separating) Galactic emission mechanisms
from the data.  Template fitting is perhaps the most basic of the
available techniques but has proven to be \emph{remarkably} powerful
despite its simplicity.  The fundamental equation to be solved is just
a linear fit of templates to the data: ${\bf w} = {\bf P}\vec{a}$
where ${\bf w}$ is a map of the WMAP data, ${\bf P}$ is a matrix of
template maps, and $\vec{a}$ is a vector of amplitudes.  The
least-squares solution to this equation (after taking the noise map
${\bf n}$ into account) is $\vec{a} = ({\bf P}^T{\bf n}^{-1}{\bf
P})^{-1}({\bf P}^T{\bf n}^{-1}{\bf w})$.  This formalism easily
accommodates partial sky coverage and arbitrary numbers of templates.
Of course, for the purpose of studying Galactic foregrounds, one of
the templates must be a CMB estimate with an input CMB spectrum, and
as shown in DF08 \citep[as well as][]{dobler08b,dobler09}, this
introduces a bias in the derived foreground spectra because no CMB
estimate is ever completely clean of the very foregrounds to be
measured.

\input{fig1}

When applying template fitting to the WMAP 7-year temperature data
using the templates and partial sky regions defined in DF08 (to take
into account foreground spectral variation with position), the 7-year
WMAP haze/bubbles emission is shown in \reffig{7yrhaze}.  However, I
have made two adjustments to the templates used in the fit compared to
DF08.  First, for the haze template, I use a bivariate Gaussian of
scale length $\sigma_{\ell}=15\degree$ and $\sigma_{b}=25\degree$.
Second, the haze residual in DF08 included a significant ``disky''
component near the mask likely associated with energy dependent
propagation lengths of electrons accelerated in the Galactic disk as
discussed in \cite{mertsch10}.  This is a failing of using the Haslam
408 MHz map \citep{haslam82} as a soft synchrotron template.
Additionally, there is the possibility that a population of pulsars in
the Galactic disk could potentially produce harder spectrum cosmic-ray
electrons.\footnote{However, the efficiency factor for generating
electrons with pulsars is presently not constrained (and could be as
low as zero), while \cite{mertsch10} showed that the energy dependent
propagation effects are of roughly the right order of magnitude given
the results in \refsec{specandmorph}.}  Whatever its origin, this
disky component likely does not represent true haze emission and as
such I include a Gaussian disk template with scale lengths
$\sigma_{\ell}=20\degree$ and $\sigma_{b}=5\degree$ .  The mask used
to fit the data excludes regions where the dust extinction at
H$\alpha$ is greater than 1 magnitude as well as all point sources in
both the WMAP and \emph{Planck} ERCSC (30 GHz to 143 GHz) catalogs,
the LMC, SMC, M31, Orion-Barnard's Loop, NGC 5128, and $\zeta$-Oph.

As seen in \reffig{7yrhaze}, outside of the haze region, the residuals
are remarkably small, indicating that the templates used in the
analysis do represent reasonable tracers of the actual
emission.\footnote{The one exception is the Gum Nebula at
$\ell\sim-100\degree$; there are two reasons for this.  First, as
pointed out in DF08 and \cite{dobler09}, the gas temperatures and
spinning dust spectra are varying very rapidly with position in this
region.  Second, the Gum Nebula is one of the brightest Galactic
regions in the raw data and so these residuals actually represent a
small fraction of the total emission.  Also, it is important to note
that the Gum Nebula is relatively nearby and as such occupies
a \emph{much} smaller volume than the haze/bubbles.}  Most
importantly, several authors have found that emission from rapidly
rotating dust grains (which have small dipole moments and thus produce
spinning dipole radiation) represents $>$30\% of the diffuse emission
at K-band \citep[see DF08 and][]{finkbeiner04a}.  This spinning dust
is perhaps the most uncertain foreground since it is not well known
how its emissivity scales with total dust column density (inferred
from 100$\mu$m for example).

In addition to the simple template fitting described above, more
involved component separation techniques have been developed and
applied to the WMAP data.  The two most commonly used are the Maximum
Entropy Method \citep{bennett03b} and Bayesian inference via Gibbs
sampling \citep{eriksen06}.  While these methods are excellent for
cleaning the CMB of foregrounds and estimating the amplitude of
emission with known morphology (such as the cosmic dipole), they are
not optimal for identifying new components.  For example, although it
has been reported that these techniques do not ``see'' the WMAP
haze/bubbles, in fact \cite{pietrobon11} show that the haze emission
is typically swept into a ``low frequency'' component that is a
conglomerate of spinning dust, soft synchrotron, hard synchrotron, and
free-free all forced to obey a single power law spectrum.

%% file: fig1.tex
\bpm
  \centerline{
    \includegraphics[width=0.49\textwidth]{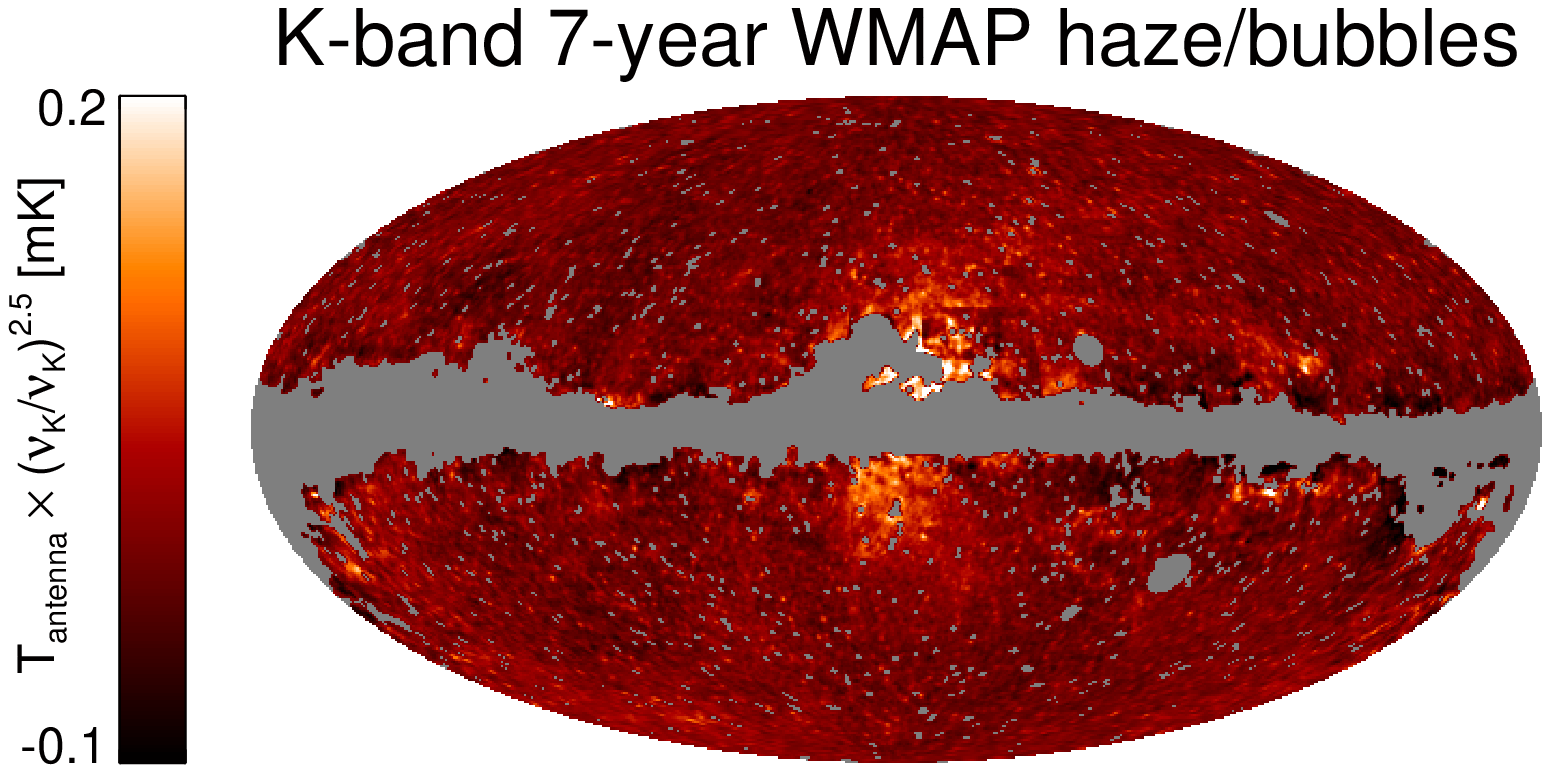}
    \includegraphics[width=0.49\textwidth]{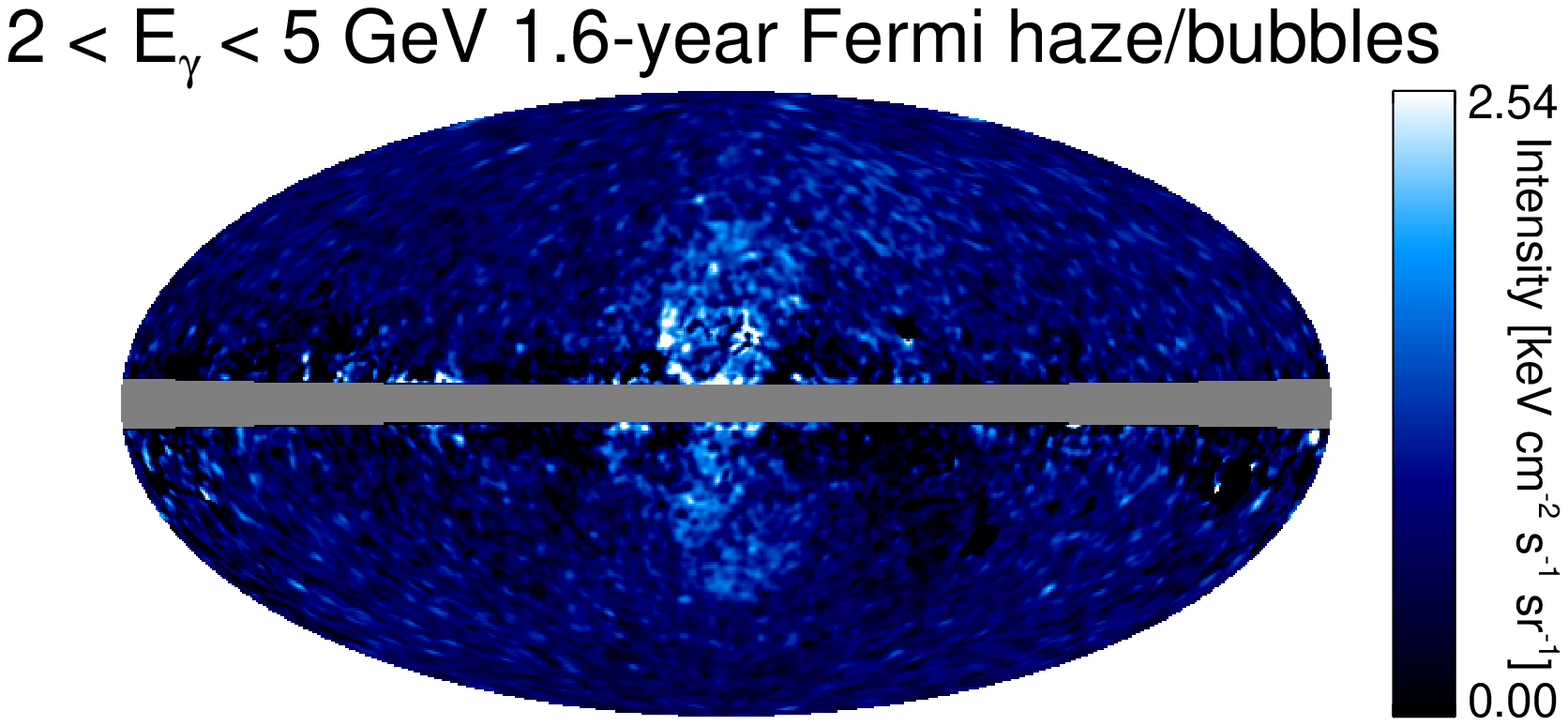}
  }
  \centerline{
    \includegraphics[width=0.49\textwidth]{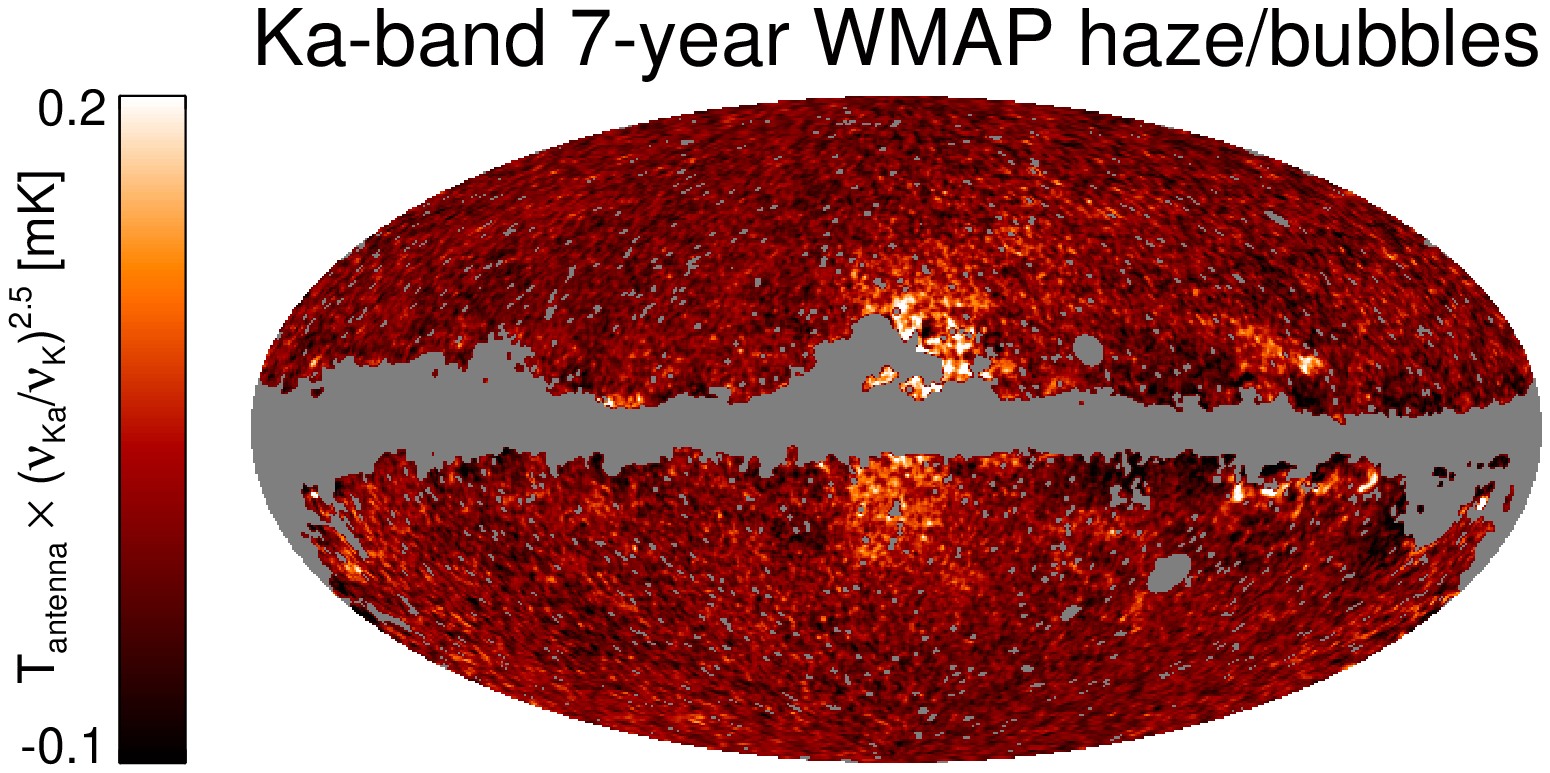}
    \includegraphics[width=0.49\textwidth]{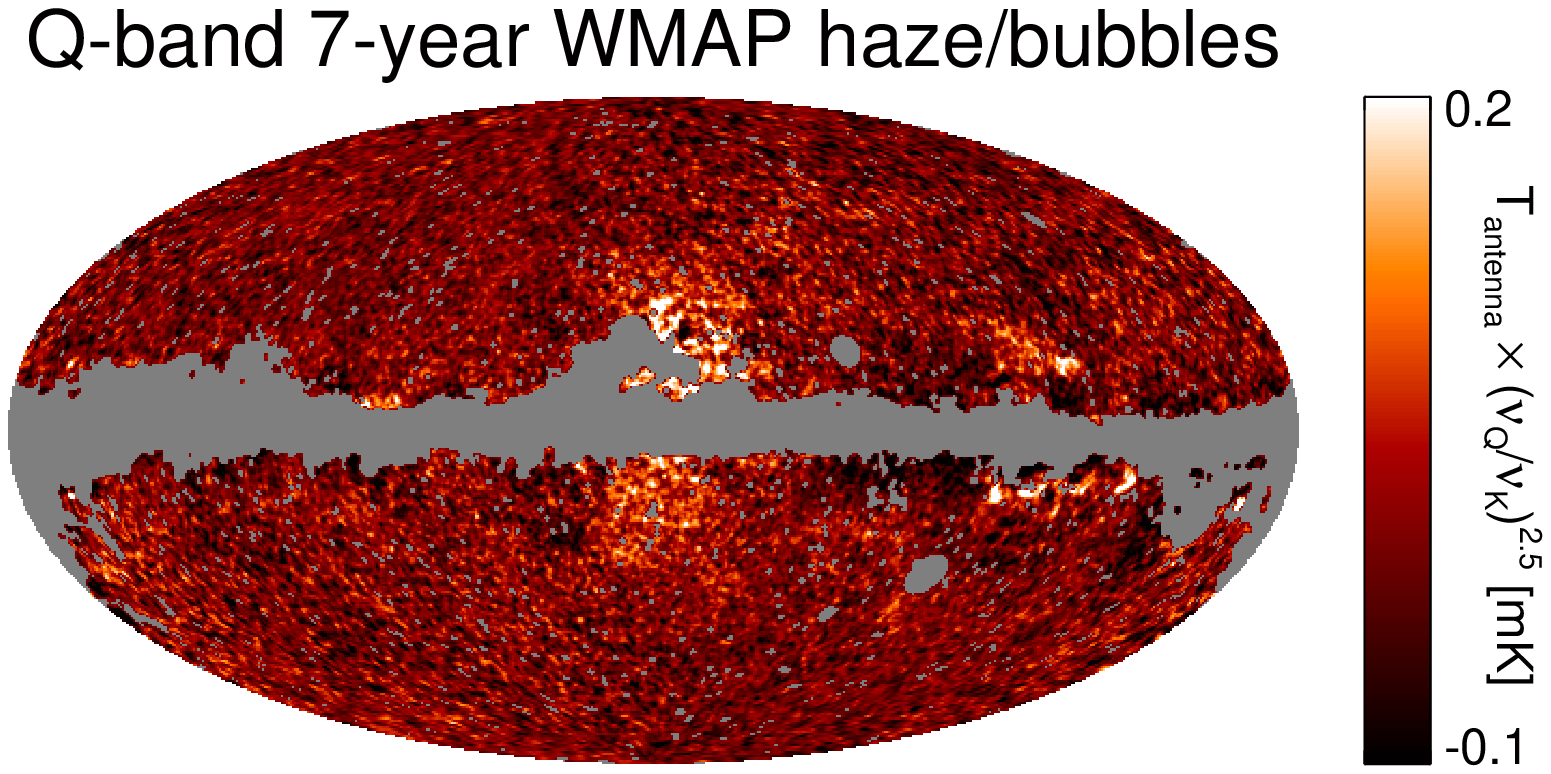}
  }
\caption{
  The haze/bubbles in both microwaves and gamma-rays (23 GHz, 33 GHz,
  41 GHz, and 2-5 GeV, counter clockwise from top left).  In the
  microwaves, templates have been used to regress out emission from
  the CMB, thermal and spinning dust, free-free, and soft synchrotron.
  In the gammas, the official \emph{Fermi} diffuse model has been
  subtracted from the data \citep[see][]{dobler10}.  In all bands, the
  haze is seen to be elongated in latitude by a factor of roughly two
  reaching $\pm50\degree$ in the gammas and $\sim\pm35\degree$ in the
  microwaves.  The microwaves are stretched with a $\nu^{2.5}$
  scaling, which yields roughly equal brightness from K to Q band,
  indicating an electron spectrum of $dN/dE_e \propto E^{-2}$, broadly
  consistent with the gamma-ray spectrum from \cite{dobler10}
  and \cite{su10}.  The gamma-ray haze/bubbles seem to have a sharp
  edge near $|b|\sim50\degree$ while the microwaves seem to fall off
  quickly for $|b|>30\degree$ (particularly in the south).
}
\label{fig:7yrhaze}
\epm

%% file: specandmorph.tex
\subsection{The Haze at Low Latitudes}

\input{fig2}

The residuals in \reffig{7yrhaze} and the spectrum plots in
\reffig{spectrum} show the two important features of the microwave
haze/bubbles.  First, the spectrum is very hard, $T_{H} \propto
\nu^{\beta_H}$ with $\beta_H \approx -2.5$ which is harder than can be
obtained from particle acceleration in SN shocks after taking into
account cosmic-ray diffusion and energy losses.  However, it is
significantly \emph{softer} than the $\beta_F = -2.15$ required if the
emission were free-free.  Second, the haze/bubbles are extended and
elongated in latitude.  In detail however, while the haze is roughly
centered on the GC, its morphology is somewhat different in the north
and south.  In the south, and above latitude $b=-30\degree$, the
morphology of the WMAP haze/bubbles is strikingly similar to the
\emph{Fermi} haze/bubbles (right panel of \reffig{7yrhaze}).  In the
north, the situation is much more complicated by the significant dusty
regions of the Ophiuchus complex.  In fact, the direction of the north
Galactic center (below $b=30\degree$) likely contains either
significant spinning dust emission that is not well approximated by
our dust template (the \cite{schlegel98} dust map evaluated at 94 GHz
by \cite{finkbeiner99}) or highly variable gas temperatures making our
free-free template (the \cite{finkbeiner03} H$\alpha$ map) unreliable.
In reality, both effects are likely operating simultaneously.

\input{fig3}

This additional complexity in the north is the reason that I
concentrate on the south when estimating the spectrum of the haze in
\reffig{spectrum}.  The derived spectrum of $\beta_H \approx -2.5$
implies an electron spectrum of $dN/dE_e \propto E_e^{-2}$ which has
been shown by \cite{dobler10} and \cite{su10} to be broadly consistent
with the spectrum of the \emph{Fermi} gamma-ray haze/bubbles.  This,
taken together with the morphological correlation at low latitudes,
provides the strongest evidence that this is in fact the same
phenomenon observed at multiple wavelengths.  While it is true that
DF08 showed that the bias induced by presubtracting a CMB estimate can
be quite large, for a component with a spectrum $T \propto \nu^{-2.5}$
the CMB5 weights for the WMAP7 data ($\zeta=[0.246, -0.736, -0.0685,
  0.263, 1.295]$ in thermodynamic $\Delta T$ for the five WMAP bands)
would lead to a measured spectrum $\propto \nu^{-2.53}$ when comparing
K- to Ka-band and $\propto \nu^{-2.54}$ when comparing K- to Q-band.
In other words, the bias for a foreground with a haze/bubbles-like
spectrum would be rather small given the DF08 CMB5 estimate and so
$\beta_H = -2.5$ is likely \emph{very} close to the true spectrum.
Furthermore, a component with a free--free spectrum would be inferred
to have a spectrum of $\propto\nu^{-2.154}$ providing more evidence
that the microwave haze/bubbles is \emph{not} due to free--free
emission.

However, it is important to bear in mind that the majority of the
haze/bubble gammas observed by \emph{Fermi} at high latitudes likely
come from $E_e > 100$ GeV electrons scattering CMB photons, while the
microwaves are generated by electrons with energies 1 GeV $< E_e <$
100 GeV.  This is illustrated in \reffig{jnuvse} which shows the total
emissivity
\be
  j_{\nu}^{\rm tot} = \int E_e j_{\nu}(E_e,B)\frac{dN}{dE_e}d\ln E_e,
\ee
where $j_{\nu}$ is the emissivity from a single electron at energy
$E_e$ in a magnetic field $B$, as a function of electron energy for a
population of electrons with the haze/bubbles spectrum in a 5$\mu$G
magnetic field.  It is also worthwhile to note that, as shown in the
figure, if the electron energy extends up to order TeV energies as
required by the gammas, then the synchrotron emissivity per electron
for $E_e>100$ GeV actually increases with increasing frequency.
However, the falling electron spectrum and emission from thermal dust
(which rises much faster with frequency) render this high frequency
synchrotron undetectable.

\subsection{The Haze at High Latitudes}

\input{fig4}

\input{fig5}

An aspect of the haze/bubbles that has not been fully discussed is the
behavior of the microwave emission at \emph{high} latitudes, where the
gamma-ray signal is most unambiguous and where the microwaves appear
to fall off in intensity.  \reffig{bigmask} shows the microwave and
gamma-ray haze side-by-side, with a large $\pm28\degree$ mask and
smoothed to a common $2\degree$ beam.  It is immediately clear from
the figure that there most certainly \emph{is} haze emission above
$|b|=28\degree$, but its morphology is somewhat different than the
gammas.

Concentrating again mostly on the south, the gammas fill a bubble with
a sharp edge while the microwaves seem to fall off dramatically for
$b<-35\degree$ as shown in the lower left panel of \reffig{bigmask}.
The intensity as a function of latitude within the bubble (defined by
the gamma-ray emission), does not quite have an ``edge'' in the same
sense as the gammas, but the fall off is quite rapid indicating either
a rapid decrease in the number of electrons or the strength of the
magnetic field.  Given the gammas at latitudes
$-50\degree<b<-35\degree$, the latter seems more likely.  Furthermore,
given that neither synchrotron maps at lower frequencies (e.g., the
408 MHz map) nor WMAP polarization show a sharp drop in synchrotron
emission below $b=-35\degree$, this seems to indicate that the
dominant field component producing the synchrotron is the field within
the haze/bubbles itself rather than the Galactic magnetic field.
Assuming that the haze/bubbles is located at the GC, then the scale
height of this magnetic field drop off is $\sim$6 kpc.

The ``sharpness'' of the microwave edge is depicted in
\reffig{hazlatlon} which shows the total intensity as a function of
both latitude and longitude in the south.  Interestingly, for
latitudes $-35\degree < b < -6\degree$, the brightness is roughly flat
with distance.  There is some slow decrease, but the drop below
$b=-35\degree$ is striking, and this flatness is reminiscent of the
gamma-ray emission --- though again, the edge is not as sharp and is
at lower absolute latitude.  This latitudinal profile is qualitatively
different from that presented in DF08 in which the haze brightness
increases sharply towards the GC above $b=-10\degree$.  The difference
is that the inclusion of the extra disk template (to account for the
excess disky emission near the mask) has removed most of this power
making the profile ``flatter'' with latitude.  As a function of
longitude, again the haze brightness is roughly constant within about
$|\ell|<15\degree$ outside of which the fall off is quite rapid.  The
figure also shows that for latitudes $-50\degree<\ell<-40\degree$,
there is \emph{no} significant microwave haze/bubbles emission, in
contrast to the gammas.

In the north, again the situation is somewhat complicated since there
is clearly emission co-located in both the microwaves and gammas, but
there is also significant contamination by spinning dust in the
microwaves and $\pi^0$ decay photons in the gammas.  Nevertheless, it
is clear from the morphology (see \reffig{bigmask}) that, while the
lower latitude microwaves are centered on the GC, the center of the
structure at $|b|=28\degree$ is roughly $\ell=-5\degree$, an offset
which causes some difficulty for formation scenarios.

%% file: fig2.tex
\bpm
  \centerline{
    \includegraphics[width=0.98\textwidth]{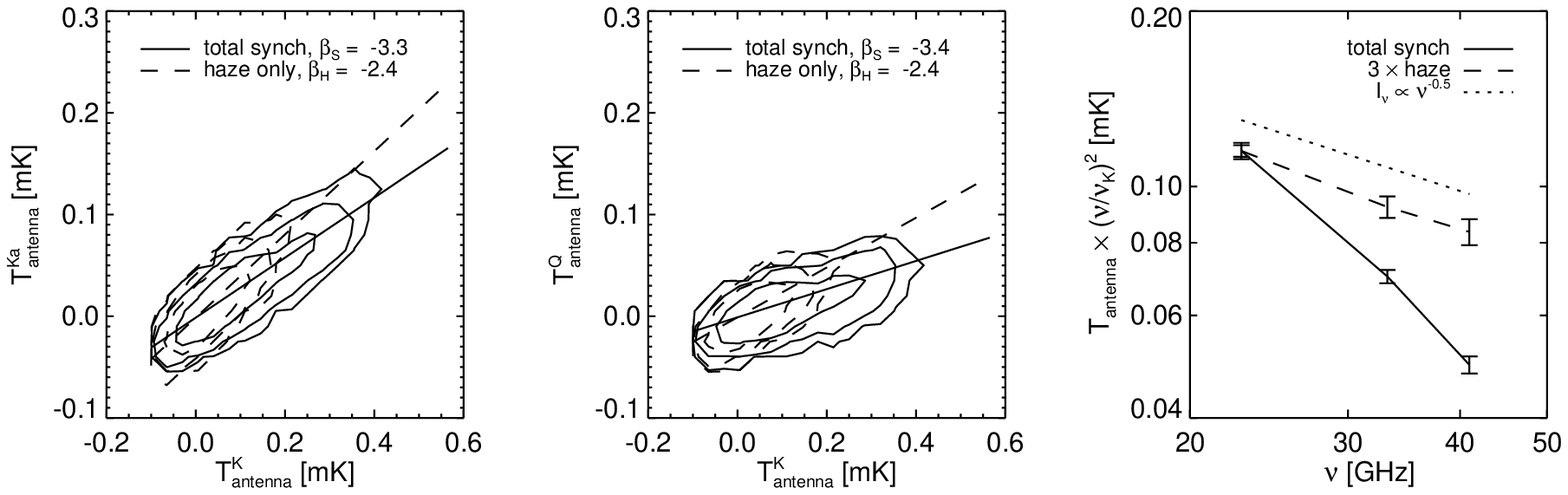}
  }
\caption{
  \emph{Left} and \emph{middle}: scatter plots (drawn with contours)
  for the microwave haze/bubbles residuals in \reffig{7yrhaze} (dashed
  lines) and for the total synchrotron (haze plus soft synchrotron;
  solid lines) in the region $|l|<25\degree$,
  $-35\degree<b<-10\degree$.  The best fit power law for this region
  shows that the haze emission is significantly \emph{harder} than the
  soft synchrotron from 23 GHz to 41 GHz.  The total emission in the
  region for both cases is shown in the \emph{right} panel.  The haze
  emission is clearly harder given the noise in the data, is
  consistent with a power law of roughly $\nu^{-2.5}$, and represents
  approximately 33\% of the total synchrotron emission at K-band.
}
\label{fig:spectrum}
\epm

%% file: fig3.tex
\bp
  \centerline{
    \includegraphics[width=0.38\textwidth]{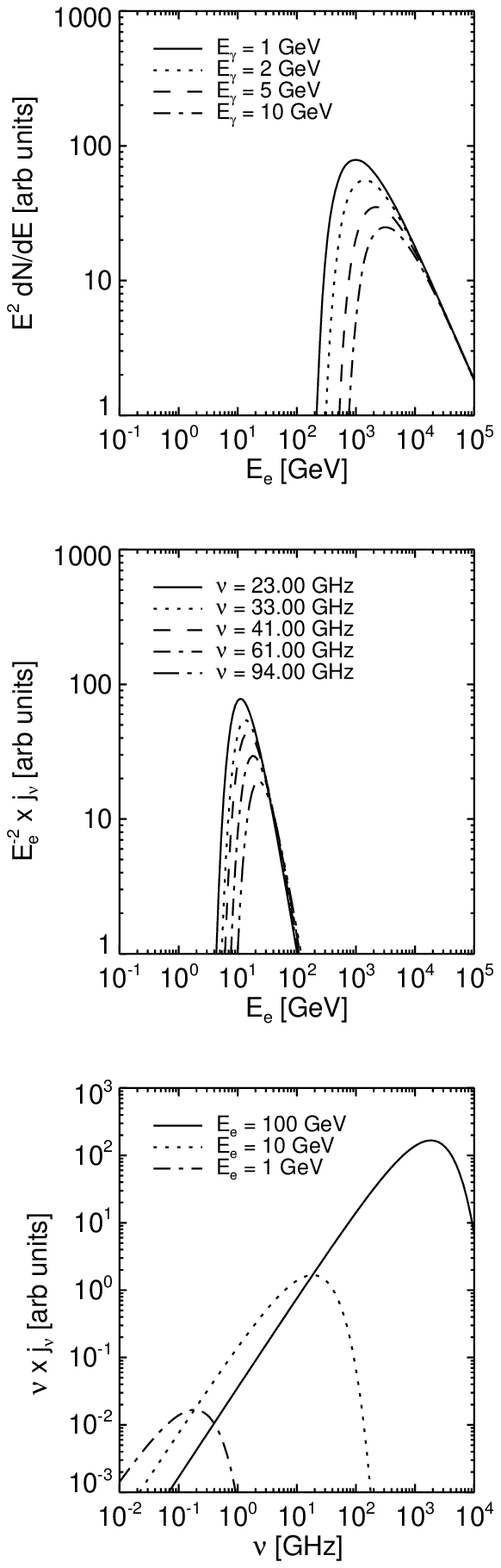}
  }
\caption{
  \emph{Top} and \emph{middle}: intensity in CMB IC scattered gammas
  and microwave synchrotron as a function of the electron energy for a
  population of electrons with spectrum $dN/dE_e \propto E^{-2}$
  illustrating that the WMAP haze/bubbles and \emph{Fermi}
  haze/bubbles are generated by different ranges of the electron
  spectrum.  In particular, if the gammas at $\pm10$ kpc are generated
  by scattering of CMB photons, this requires $\sim$TeV electrons at
  those distances, while the microwaves are generated by electrons
  with energies $\sim$10 GeV.  The higher energy electrons used to
  generate the gammas would create a synchrotron signal that peaks at
  very high frequencies (\emph{bottom}), though it would be
  overshadowed by thermal dust emission.
}
\label{fig:jnuvse}
\ep

%% file: fig4.tex
\bpm
\centerline{
    \includegraphics[width=0.70\textwidth]{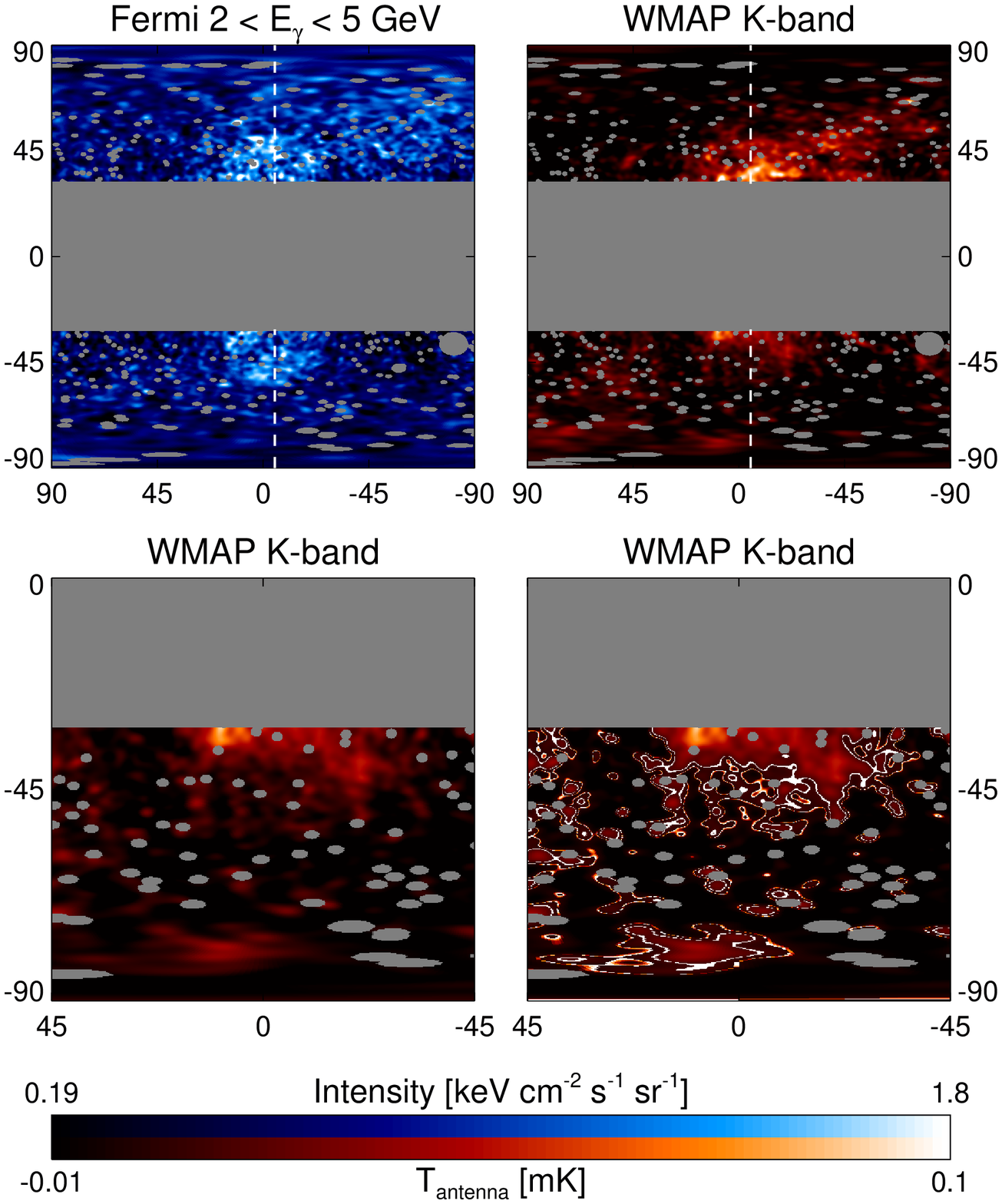}
}
\caption{
  The \emph{Fermi} (\emph{top left}) and WMAP (\emph{top right})
  haze/bubbles at high latitudes masking emission for $|b|<28\degree$.
  The correlation between the two is clear, though the microwave
  ``edge'' is at a lower latitude than the gamma-ray edge, indicating
  that the ``bubble'' is not completely filled by microwaves.  The
  dashed white line at $\ell=-5\degree$ illustrates that, while the
  haze/bubbles is centered on the GC at low latitudes, it is offset by
  a few degrees above $|b|\sim28\degree$.  The sharp drop in
  microwaves in the Galactic south (\emph{bottom left}) indicates
  either a sharp decrease in the total number of electrons or a sharp
  drop in the magnetic field.  The former seems unlikely given the
  presence of gammas down to $b\sim-50\degree$.  \emph{Bottom right}
  the same stretch, but with contours showing $T_{\rm antenna} =
  10^{-2}$ and $10^{-3}$ mK indicating the very sharp transition from
  haze/bubbles to noise below $b\sim-30\degree$ (see
  also \reffig{hazlatlon}).
}
\label{fig:bigmask}
\epm

%% file: fig5.tex
\bpm
\centerline{
    \includegraphics[width=0.98\textwidth]{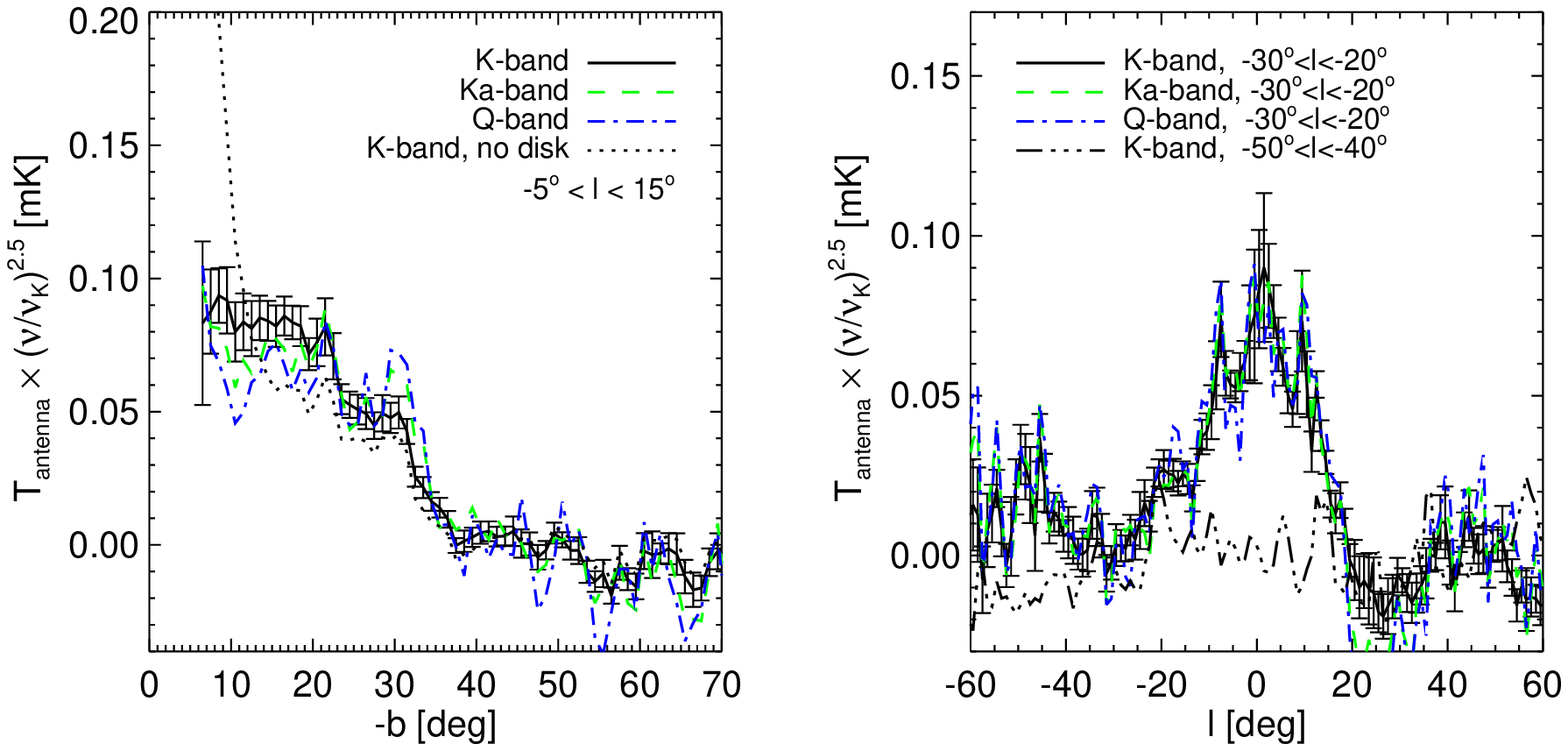}
}
\caption{
  The microwave haze/bubbles profile as a function of latitude
  (\emph{left}) and longitude (\emph{right}).  Inside of the haze
  region, the brightness is roughly flat (or at most, decreases slowly
  with latitude) and then drops sharply for latitudes below $b \approx
  -35\degree$ and for longitudes $|\ell|>15\degree$.  This ``edge'' in
  the microwaves is not as sharp as in the gammas, and the dotted line
  in the right panel indicates that there is little to no microwave
  emission for latitudes $-50\degree < b < -40\degree$, in contrast
  with the gamma-ray measurements, indicating a relatively quick drop
  in the magnetic field at $b\sim-35\degree$.  The dotted line in the
  left panel shows the results without including the bivariate
  Gaussian disk template.  In that case, the profile rises
  dramatically towards the GC, but this emission is not likely to be
  associated with the haze (see \refsec{methods}).
}
\label{fig:hazlatlon}
\epm

%% file: polarization.tex
\input{fig6}

Given the spectrum derived in \reffig{spectrum}, the haze emission
certainly seems to be consistent with hard spectrum synchrotron.  Also
given that synchrotron with $\beta\approx-2.5$ can be up to 70\%
polarized in a completely ordered magnetic field, this raises the
question whether the haze appears in WMAP polarization data.  Several
authors have addressed this issue, most notably \cite{gold11} who
claim no evidence for a significant hard component from WMAP 7-year
data.  However, this claim suffers from two major drawbacks.  First,
as shown in \cite{miville-deschenes08} and argued in DF08, even a
small turbulent component in the magnetic field can reduce the
polarization amplitude when projecting along the line of sight.  In
fact, this line of sight depolarization is clearly observed in the
WMAP data as illustrated in \reffig{polscale} which shows that
(outside of the mask used in the fits), almost \emph{all} of the
emission is due to Loop I which is a local feature.  The disk
synchrotron is simply not present in the polarization data though it
is seen in the total intensity data, indicating that most of the disk
emission is line-of-sight depolarized by the varying orientation of
the Galactic magnetic field.

As \reffig{polscale} illustrates, the second shortcoming is that, even
if the polarization fraction of the haze were constant with respect to
the polarization fraction of prominent features like the Loop I SN
remnant, the noise in the polarization data is prohibitively large to
detect a harder synchrotron component.  The argument goes as follows,
forming the polarization residual
\be
  R_{P} = P_{\rm Ka} - P_{\rm K}\times\left(\frac{\nu_{\rm
          Ka}}{\nu_{\rm K}}\right)^{-3.3},
\ee
where $P_i$ and $\nu_i$ are the WMAP polarization map and frequency at
band $i$ respectively, produces a map consistent with noise as shown
in \reffig{polscale}.  That is, all of the emission (synchrotron in
the case of polarization) outside of the mask is consistent with
having a single power law spectrum $\beta_S = -3.3$.  However, if we
make the same residual using just the synchrotron intensity $S_i$,
\be
  R_{I} = S_{\rm Ka} - S_{\rm K}\times\left(\frac{\nu_{\rm
          Ka}}{\nu_{\rm K}}\right)^{-3.3},
\ee
where $S_i$ is the template extracted synchrotron (i.e., the raw data
minus the CMB, thermal plus spinning dust, and free-free
components), \reffig{polscale} shows that this residual \emph{also}
shows no evidence of the haze.  However, we \emph{know} that the haze
is there and that its spectrum is harder than elsewhere in the Galaxy
from Figures \ref{fig:7yrhaze}-\ref{fig:bigmask} and Figure 7 of DF08.
The conclusion is that, given the short lever arm of 23-33 GHz, even
the temperature data are too noisy to extract a slightly harder
spectrum component in this manner, let alone the polarization data.

%% file: fig6.tex
\bpm
  \centerline{
    \includegraphics[width=0.49\textwidth]{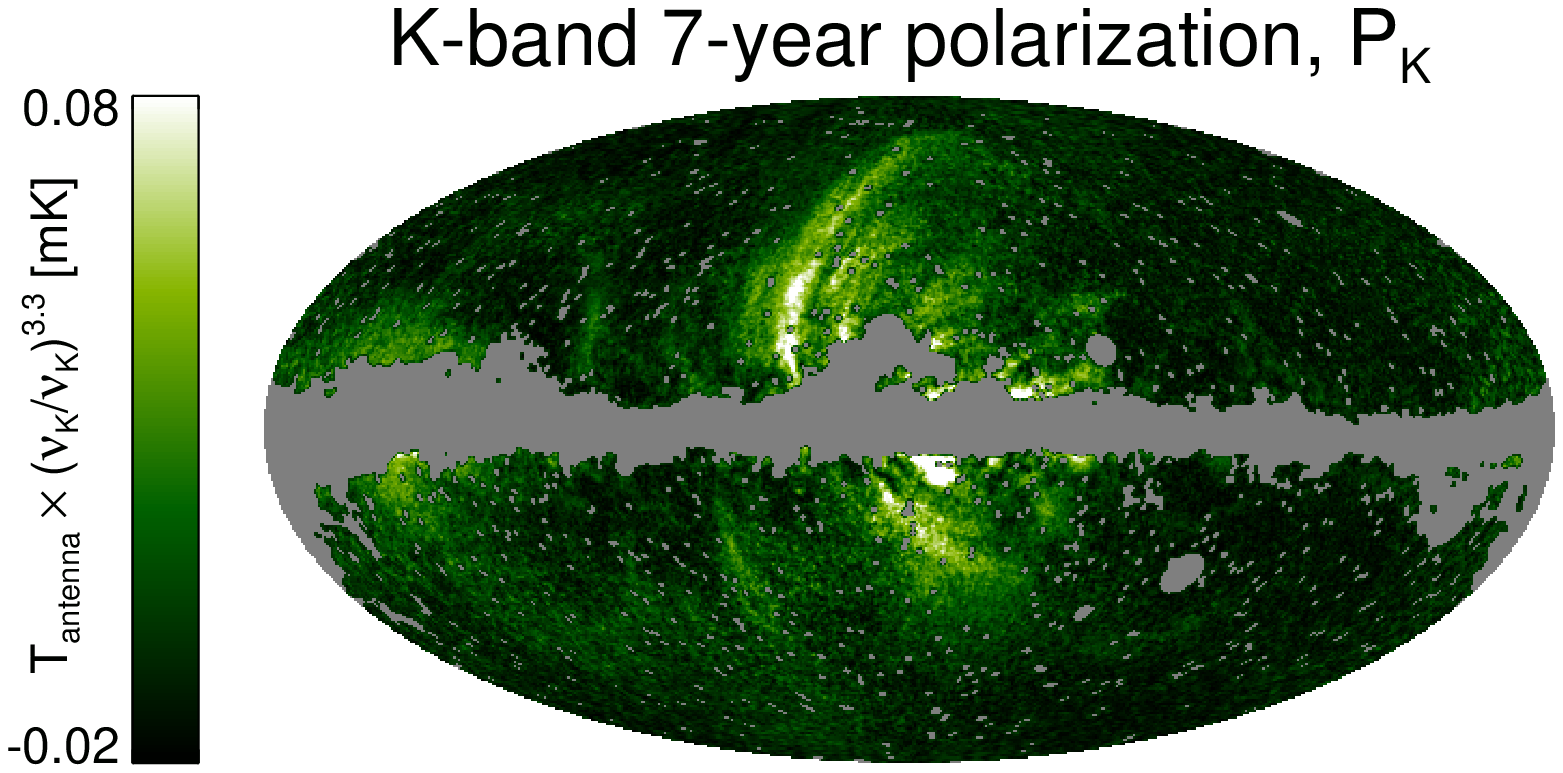}
    \includegraphics[width=0.49\textwidth]{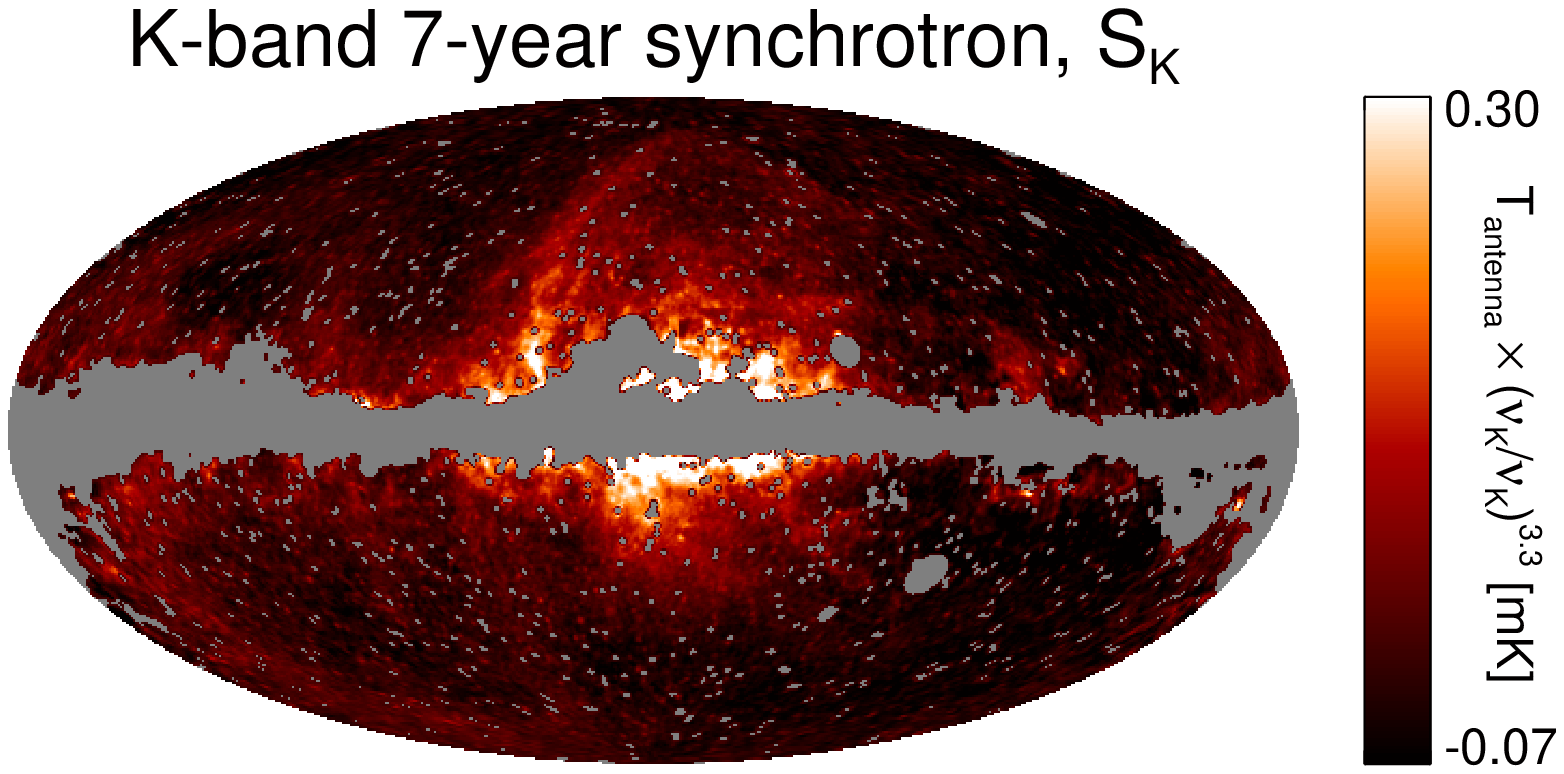}
  }
  \centerline{
    \includegraphics[width=0.49\textwidth]{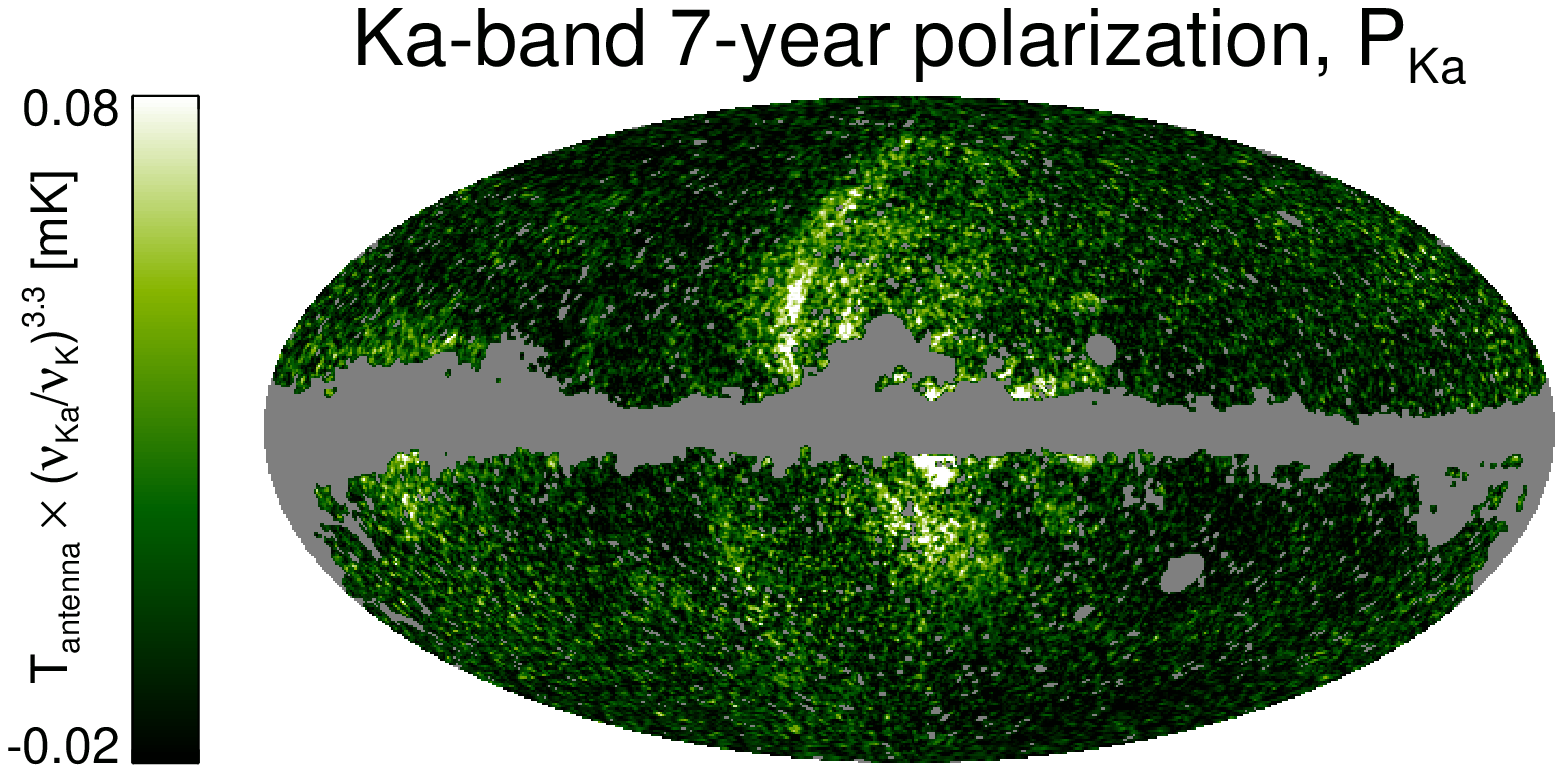}
    \includegraphics[width=0.49\textwidth]{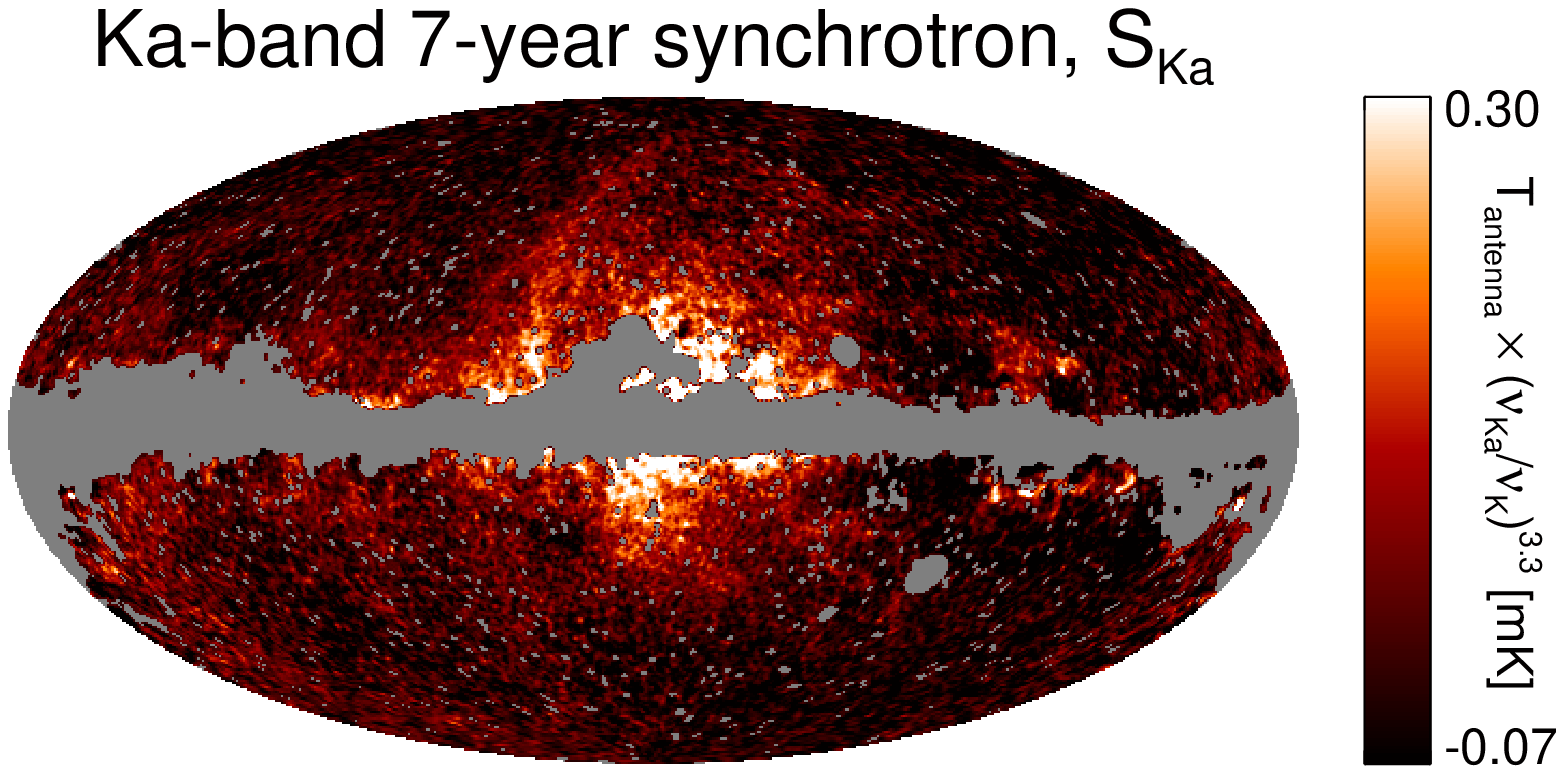}
  }
  \centerline{
    \includegraphics[width=0.49\textwidth]{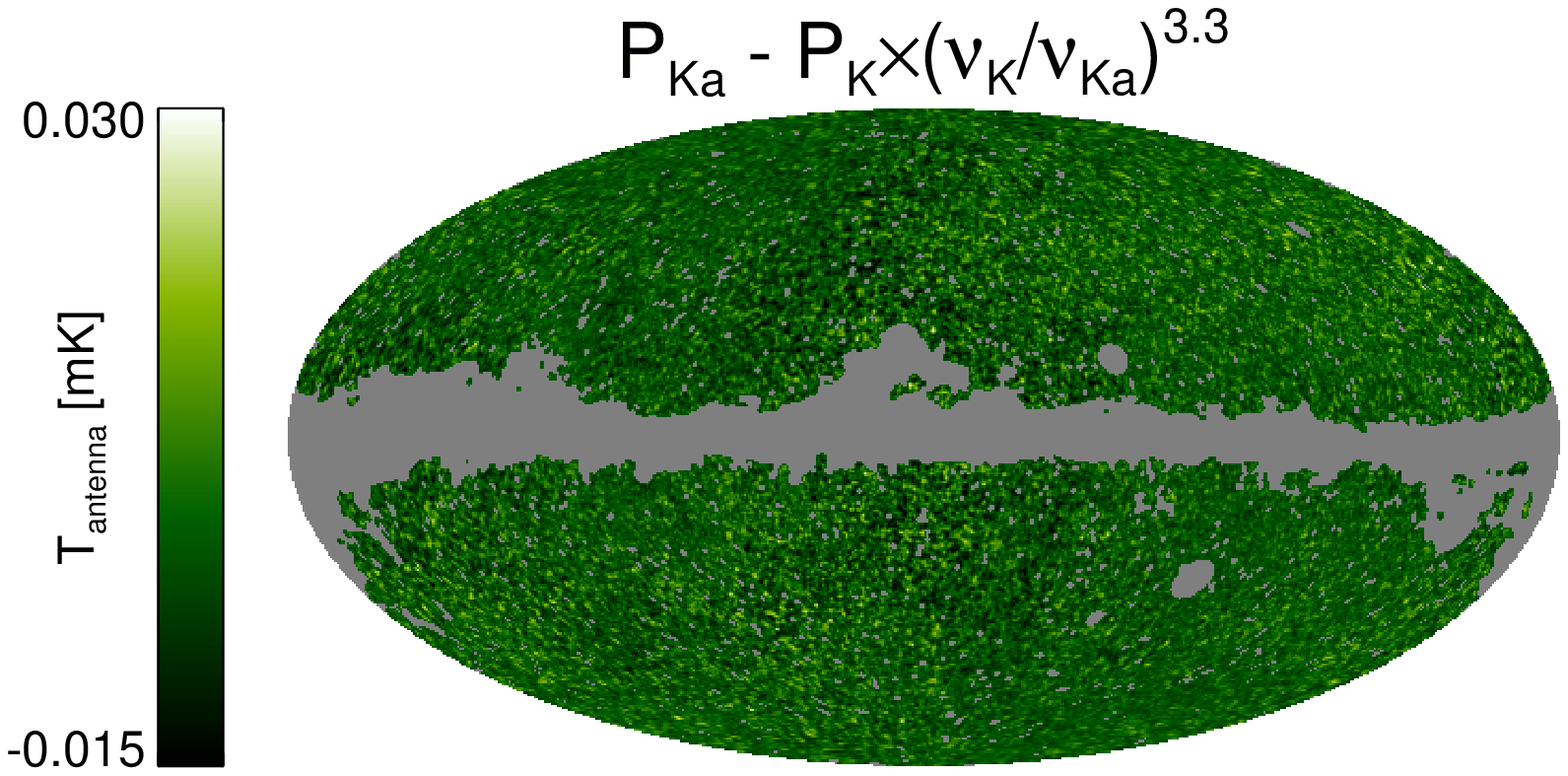}
    \includegraphics[width=0.49\textwidth]{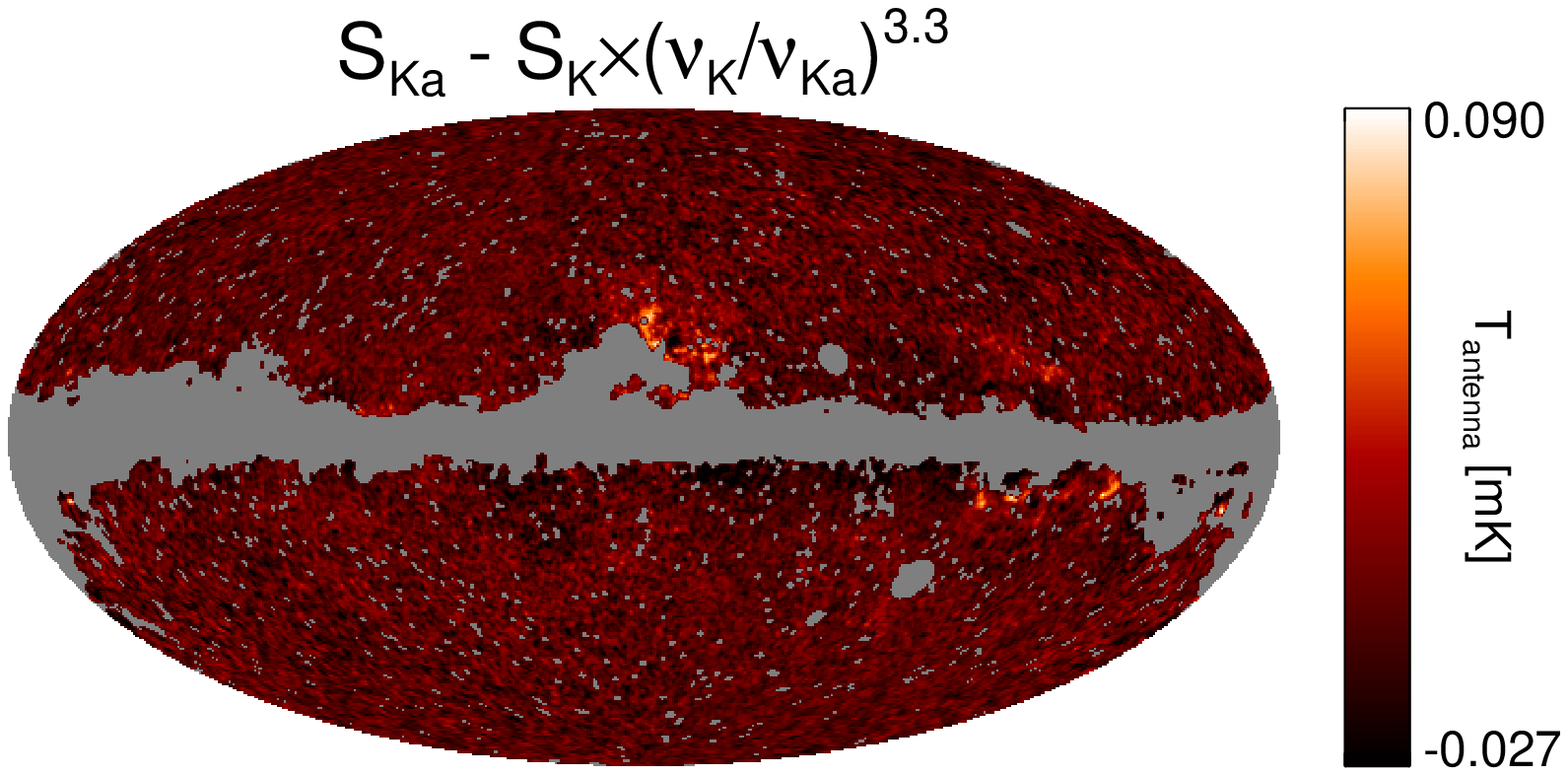}
  }
\caption{
  The WMAP polarization (\emph{left}) and total synchrotron intensity
  (\emph{right}) at K- and Ka-band (\emph{top two rows}).  Although
  some features are similar (namely Loop I in the North), there are
  significant regions of synchrotron intensity that do not appear in
  polarization (e.g., the Galactic plane) indicating line-of-sight
  projection effects through a turbulent magnetic field that reduce
  the total polarized signal.  Scaling $P_{\rm K}$ by a single
  $\beta=-3.3$ power law across the whole sky and then subtracting
  from $P_{\rm Ka}$ yields residuals consistent with noise
  (\emph{bottom left panel}) potentially indicating that the haze does
  not appear in polarization.  However, a similar exercise with the
  total intensity also yields similar results (\emph{bottom right
  panel}) \emph{despite} the fact that the haze/bubbles are there and
  have a different spectrum (see Figures \ref{fig:7yrhaze}
  and \ref{fig:spectrum}).  The implication is that the 23-33 GHz
  lever arm is too small and the noise in the polarization data too
  large to positively identify or rule out a polarized haze/bubbles
  component.
}
\label{fig:polscale}
\epm

%% file: interpretation.tex
Since the haze/bubbles was first discovered in the microwaves by
\cite{finkbeiner04a}, and especially since the recent discovery in the
gammas by \cite{dobler10}, there has been significant effort devoted
to theorizing about the origin of this hard spectrum population of
electrons.  Each scenario has associated pros and cons, successfully
producing some aspects of the emission and failing to produce others.
The following is a list of the leading possibilities put forth in the
literature.

{\bf Galactic wind:} The Galactic wind scenario 
\citep{crocker11a,crocker11b} suggests that cosmic-rays are
accelerated to $\sim$TeV energies, that the gammas are the result of
collisions of cosmic-ray protons with a very underdense ISM in the
bubbles producing $\pi^0$ decay emission, and that the microwaves
represent the synchrotron from secondary electrons.  This model can
reproduce the hard spectrum WMAP signal and, because the proton
spectrum is also very hard, the hard spectrum gamma-rays as well.  The
primary failing of this model is that it requires extended injection
of cosmic-rays over several billions of years, making the sharp edge
of the gammas and microwaves very difficult to maintain over those
long time scales.  Additionally, even in the event where the ISM is
``saturated'' with protons as they describe, the bubbles will produce
constant volume emissivity leading to limb darkening of the gammas
contrary to observations.  Finally, as in observed winds in other
galaxies \citep{veilleux05}, we would expect to see a significant cool
component producing H$\alpha$ emission, but no H$\alpha$ associated
with the haze/bubbles has yet been observed \citep[DF08;][]{su10}.

{\bf Starburst/SNe:} In this scenario a significant outflow is
generated by star formation and/or SNe in the
GC \citep[e.g.,][]{biermann10}.  However, this model would not only
likely lead to associated H$\alpha$, but it is not clear how the
uniform intensity gammas are produced or the sharp edges.
Furthermore, the timescale for the cosmic-rays to reach 5-10 kpc as
required by the microwave and gamma-ray data implies that the spectrum
would be rendered too soft to reproduce the haze/bubbles given that
the energy loss time for $\sim100$ GeV electrons is
$\tau\sim10^5-10^6$ yr.  Reacceleration of the cosmic-rays is a
possibility, but what the reacceleration mechanism is above a few kpc
is unclear.

{\bf Second order \emph{Fermi} acceleration:} Rather than an actual
``formation'' mechanism, this model relies on the haze/bubbles to be
generated by, for example, a GC jet (see below) which is then filled
with magnetosonic turbulent waves that accelerate
particles \citep{mertsch11}.  At the source of injection, the electron
spectrum is $dN/dE_e \propto E_e^{-2}$.  This scenario suffers from
having relatively little predictive power since the absolute number of
accelerated electrons relies on the microphysics of the acceleration
mechanism (which is not well known) and the injection morphology can
be set somewhat arbitrarily (though it must be concentrated near the
outer shock).  Furthermore, as found by \cite{mertsch11}, for the
required parameters to explain the gammas, the WMAP haze is
underpredicted by an order of magnitude.  This tension can be
alleviated somewhat by increasing the magnetic field strength, but
this would predict microwaves below $b=-35\degree$ (and indeed a
bubble edge in the microwaves at $|b|\sim50\degree$) which is not seen
in the data.

{\bf Active Galactic nucleus:} One of the most promising scenarios is
that there was some accretion event onto the central Galactic black
hole roughly 1 million years ago that resulted in a GC
jet \citep{guo11a}.  This model is attractive mainly because it is
episodic (meaning that it can more easily explain the sharp edges in
the gammas) and it can produce roughly the required integrated power
in gammas and microwaves.  There are four primary concerns with the
model however.  First, as shown by \cite{su10} the intensity profile
of the gammas is roughly uniform and, at best, the AGN model would
predict a uniform volume emissivity, again yielding limb darkening,
especially near the base of the jet generated bubble which is very
thin.  Second, generically, shearing instabilities are generated at
the edge which are not seen in the gamma ray data which has smooth
edges.  However, viscosity within the plasma can effectively suppress
the instabilities \citep{guo11b}.  Additionally, magnetic draping may
suppress the instabilities (and have the added benefit of confining
the cosmic-rays as required), though it has not yet been shown that
the required field strengths would not produce synchrotron at
latitudes $|b|>35\degree$ which is not observed.  Relatedly, the third
concern is that, from jets in other galaxies, we see
radio \emph{lobes} -- large bubbles of radio emission at high
latitudes \citep[e.g.,][]{schreier79}.  As shown in \reffig{7yrhaze}
however, the WMAP haze is confined to \emph{lower} latitudes and large
radio lobes coincident with the gamma-rays are not present.  Finally,
while the integrated power is sufficient to explain the haze, the
model has the significant disadvantage that it has yet to predict a
spectrum for the accelerated electrons, which is a key characteristic
of the haze.  The softening argument from above still applies here and
the requirements are that the spectrum 10 kpc away from the event have
$dN/dE_e \propto E_e^{-2}$.

{\bf Dark matter annihilation:} In this model, the dark halo of the
Milky Way is composed of particles which have a self-annihilation
cross-section and number density sufficient to explain the required
injected power.  This scenario was first explored in the microwaves
by \cite{finkbeiner04b} and \cite{hooper07} and then expanded to
include local cosmic-ray measurements by \cite{cholis09b}.
Recently, \cite{dobler11} showed that the gamma-ray spectrum,
amplitude, and morphology can also be reproduced with dark matter
annihilation if the dark halo is prolate and diffusion occurs
preferentially along ordered field lines in the GC.  Here the one
significant failing of this model is that it does not produce sharp
edges as seen in the \emph{Fermi} data (the microwave morphology can
be completely dominated by the magnetic field geometry).  Thus, if
these sharp edges persist with future data, the dark matter
annihilation only model will be disfavored.

Of course, it is important to bear in mind that there may be multiple
mechanisms operating at once (e.g., a bubble blown by a jet and filled
with electrons by dark matter annihilation or a GC wind).  In fact,
this may be the only way to reconcile the very unusual features seen
in the WMAP and \emph{Fermi} data, especially given the flat
brightness profile of the haze/bubbles which is not reproduced in any
of the above models.

%% file: summary.tex
I have presented a last look at the WMAP haze/bubbles from the WMAP
7-year data, prior to the \emph{Planck} data release.  The haze
morphology and spectrum are similar to previous analyses with several
important differences.  The inclusion of a disk template in the
regression analysis suggests that the haze profile is somewhat flatter
with latitude than previously thought.  Furthermore, the sensitivity
with seven years of data does permit a study of the emission at high
(southern) latitudes.  Here I find that there is haze emission at
latitudes $|b|>28\degree$ and that there is a relatively sharp fall
off in intensity for $b<-35\degree$, most likely indicating a rapid
fall off in the magnetic field strength at $\sim$6 kpc within the
haze/bubbles itself.  As previously reported, the spectrum of the WMAP
haze is too soft to be free-free emission and too hard to be generated
by SN shocks (the ``normal'' particle acceleration mechanism) given
diffusion effects.  The spectrum is consistent with that required to
produce the \emph{Fermi} gamma-ray haze/bubbles, though I have
illustrated that the two emissions at high latitudes are likely coming
from mostly distinct parts of the cosmic-ray spectrum.  While the haze
is not strongly seen in the 7-year WMAP polarization data, I have
shown that, even if the haze were not depolarized by turbulence in the
magnetic field, the noise in the WMAP polarization data is sufficient
to render a hard spectrum polarized component undetectable by
comparison of K-band and Ka-band.

Lastly, everything that we have discovered to date about the
haze/bubbles (hard spectrum synchrotron and gammas, sharp edges in the
gammas, $\pm10$ kpc extent of the electrons, microwaves confined to
lower latitudes, uniform intensity in gammas, lack of strong
associated H$\alpha$ and strong polarization, low energy cutoff in
electron spectrum, etc.) has made pinpointing a single underlying
origin for the electron population very difficult.  All of the
proposed mechanisms have associated problems and if all of the
haze/bubbles characteristics persist with future data, hybrid
formation scenarios will likely be required.

%% file: acknowledgements.tex
\vskip 0.15in {\bf \noindent Acknowledgments:} I thank Tracy Slatyer,
Ilias Cholis, Neal Weiner, Fulai Guo, Peng Oh, Doug Finkbeiner, and
Igor Moskalenko for useful conversations.  This work has been
supported by the Harvey L.\ Karp Discovery Award.

%% file: haze7.bbl
\begin{thebibliography}{31}
\expandafter\ifx\csname natexlab\endcsname\relax\def\natexlab#1{#1}\fi

\bibitem[{{Bennett} {et~al.}(2003{\natexlab{a}}){Bennett}, {Hill}, {Hinshaw},
  {Nolta}, {Odegard}, {Page}, {Spergel}, {Weiland}, {Wright}, {Halpern},
  {Jarosik}, {Kogut}, {Limon}, {Meyer}, {Tucker}, \& {Wollack}}]{bennett03b}
{Bennett}, C.~L., {Hill}, R.~S., {Hinshaw}, G., {et~al.} 2003{\natexlab{a}},
  \apjs, 148, 97

\bibitem[{{Bennett} {et~al.}(2003{\natexlab{b}}){Bennett}, {Halpern},
  {Hinshaw}, {Jarosik}, {Kogut}, {Limon}, {Meyer}, {Page}, {Spergel}, {Tucker},
  {Wollack}, {Wright}, {Barnes}, {Greason}, {Hill}, {Komatsu}, {Nolta},
  {Odegard}, {Peiris}, {Verde}, \& {Weiland}}]{bennett03a}
{Bennett}, C.~L., {Halpern}, M., {Hinshaw}, G., {et~al.} 2003{\natexlab{b}},
  \apjs, 148, 1

\bibitem[{{Biermann} {et~al.}(2010){Biermann}, {Becker}, {Caceres}, {Meli},
  {Seo}, \& {Stanev}}]{biermann10}
{Biermann}, P.~L., {Becker}, J.~K., {Caceres}, G., {et~al.} 2010, \apjl, 710,
  L53

\bibitem[{{Cholis} {et~al.}(2009{\natexlab{a}}){Cholis}, {Dobler},
  {Finkbeiner}, {Goodenough}, {Slatyer}, \& {Weiner}}]{cholis09a}
{Cholis}, I., {Dobler}, G., {Finkbeiner}, D.~P., {et~al.} 2009{\natexlab{a}},
  arXiv:0907.3953

\bibitem[{{Cholis} {et~al.}(2009{\natexlab{b}}){Cholis}, {Dobler},
  {Finkbeiner}, {Goodenough}, \& {Weiner}}]{cholis09b}
{Cholis}, I., {Dobler}, G., {Finkbeiner}, D.~P., {Goodenough}, L., \& {Weiner},
  N. 2009{\natexlab{b}}, \prd, 80, 123518

\bibitem[{{Crocker} \& {Aharonian}(2011)}]{crocker11b}
{Crocker}, R.~M., \& {Aharonian}, F. 2011, Physical Review Letters, 106, 101102

\bibitem[{{Crocker} {et~al.}(2011){Crocker}, {Jones}, {Aharonian}, {Law},
  {Melia}, \& {Ott}}]{crocker11a}
{Crocker}, R.~M., {Jones}, D.~I., {Aharonian}, F., {et~al.} 2011, \mnras, 411,
  L11

\bibitem[{{Dobler} {et~al.}(2011){Dobler}, {Cholis}, \& {Weiner}}]{dobler11}
{Dobler}, G., {Cholis}, I., \& {Weiner}, N. 2011, \apj, 741, 25

\bibitem[{{Dobler} {et~al.}(2009){Dobler}, {Draine}, \&
  {Finkbeiner}}]{dobler09}
{Dobler}, G., {Draine}, B., \& {Finkbeiner}, D.~P. 2009, \apj, 699, 1374

\bibitem[{{Dobler} \& {Finkbeiner}(2008{\natexlab{a}})}]{dobler08a}
{Dobler}, G., \& {Finkbeiner}, D.~P. 2008{\natexlab{a}}, \apj, 680, 1222

\bibitem[{{Dobler} \& {Finkbeiner}(2008{\natexlab{b}})}]{dobler08b}
---. 2008{\natexlab{b}}, \apj, 680, 1235

\bibitem[{{Dobler} {et~al.}(2010){Dobler}, {Finkbeiner}, {Cholis}, {Slatyer},
  \& {Weiner}}]{dobler10}
{Dobler}, G., {Finkbeiner}, D.~P., {Cholis}, I., {Slatyer}, T., \& {Weiner}, N.
  2010, \apj, 717, 825

\bibitem[{{Eriksen} {et~al.}(2006){Eriksen}, {Dickinson}, {Lawrence},
  {Baccigalupi}, {Banday}, {G{\'o}rski}, {Hansen}, {Lilje}, {Pierpaoli},
  {Seiffert}, {Smith}, \& {Vanderlinde}}]{eriksen06}
{Eriksen}, H.~K., {Dickinson}, C., {Lawrence}, C.~R., {et~al.} 2006, \apj, 641,
  665

\bibitem[{{Finkbeiner}(2003)}]{finkbeiner03}
{Finkbeiner}, D.~P. 2003, \apjs, 146, 407

\bibitem[{{Finkbeiner}(2004{\natexlab{a}})}]{finkbeiner04a}
---. 2004{\natexlab{a}}, \apj, 614, 186

\bibitem[{{Finkbeiner}(2004{\natexlab{b}})}]{finkbeiner04b}
---. 2004{\natexlab{b}}, arXiv:astro-ph/0409027

\bibitem[{{Finkbeiner} {et~al.}(1999){Finkbeiner}, {Davis}, \&
  {Schlegel}}]{finkbeiner99}
{Finkbeiner}, D.~P., {Davis}, M., \& {Schlegel}, D.~J. 1999, \apj, 524, 867

\bibitem[{{Gold} {et~al.}(2011){Gold}, {Odegard}, {Weiland}, {Hill}, {Kogut},
  {Bennett}, {Hinshaw}, {Chen}, {Dunkley}, {Halpern}, {Jarosik}, {Komatsu},
  {Larson}, {Limon}, {Meyer}, {Nolta}, {Page}, {Smith}, {Spergel}, {Tucker},
  {Wollack}, \& {Wright}}]{gold11}
{Gold}, B., {Odegard}, N., {Weiland}, J.~L., {et~al.} 2011, \apjs, 192, 15

\bibitem[{{Guo} \& {Mathews}(2011)}]{guo11a}
{Guo}, F., \& {Mathews}, W.~G. 2011, submitted to ApJ, arXiv:1103.0055

\bibitem[{{Guo} {et~al.}(2011){Guo}, {Mathews}, {Dobler}, \& {Oh}}]{guo11b}
{Guo}, F., {Mathews}, W.~G., {Dobler}, G., \& {Oh}, S.~P. 2011, submitted to
  ApJ, arXiv:1110.0834

\bibitem[{{Haslam} {et~al.}(1982){Haslam}, {Salter}, {Stoffel}, \&
  {Wilson}}]{haslam82}
{Haslam}, C.~G.~T., {Salter}, C.~J., {Stoffel}, H., \& {Wilson}, W.~E. 1982,
  \aaps, 47, 1

\bibitem[{{Hooper} {et~al.}(2007){Hooper}, {Finkbeiner}, \&
  {Dobler}}]{hooper07}
{Hooper}, D., {Finkbeiner}, D.~P., \& {Dobler}, G. 2007, \prd, 76, 083012

\bibitem[{{Mertsch} \& {Sarkar}(2010)}]{mertsch10}
{Mertsch}, P., \& {Sarkar}, S. 2010, JCAP, 10, 19

\bibitem[{{Mertsch} \& {Sarkar}(2011)}]{mertsch11}
---. 2011, Physical Review Letters, 107, 091101

\bibitem[{{Miville-Desch{\^e}nes} {et~al.}(2008){Miville-Desch{\^e}nes},
  {Ysard}, {Lavabre}, {Ponthieu}, {Mac{\'{\i}}as-P{\'e}rez}, {Aumont}, \&
  {Bernard}}]{miville-deschenes08}
{Miville-Desch{\^e}nes}, M.-A., {Ysard}, N., {Lavabre}, A., {et~al.} 2008,
  \aap, 490, 1093

\bibitem[{{Pietrobon} {et~al.}(2011){Pietrobon}, {Gorski}, {Bartlett},
  {Banday}, {Dobler}, {Colombo}, {Pagano}, {Rocha}, {Saha}, {Jewell},
  {Hildebrandt}, {Eriksen}, \& {Lawrence}}]{pietrobon11}
{Pietrobon}, D., {Gorski}, K.~M., {Bartlett}, J., {et~al.} 2011, submitted to
  ApJ, arXiv:1110.5418

\bibitem[{{Schlegel} {et~al.}(1998){Schlegel}, {Finkbeiner}, \&
  {Davis}}]{schlegel98}
{Schlegel}, D.~J., {Finkbeiner}, D.~P., \& {Davis}, M. 1998, \apj, 500, 525

\bibitem[{{Schreier} {et~al.}(1979){Schreier}, {Feigelson}, {Delvaille},
  {Giacconi}, {Grindlay}, {Schwartz}, \& {Fabian}}]{schreier79}
{Schreier}, E.~J., {Feigelson}, E., {Delvaille}, J., {et~al.} 1979, \apjl, 234,
  L39

\bibitem[{{Spergel} {et~al.}(2003){Spergel}, {Verde}, {Peiris}, {Komatsu},
  {Nolta}, {Bennett}, {Halpern}, {Hinshaw}, {Jarosik}, {Kogut}, {Limon},
  {Meyer}, {Page}, {Tucker}, {Weiland}, {Wollack}, \& {Wright}}]{spergel03}
{Spergel}, D.~N., {Verde}, L., {Peiris}, H.~V., {et~al.} 2003, \apjs, 148, 175

\bibitem[{{Su} {et~al.}(2010){Su}, {Slatyer}, \& {Finkbeiner}}]{su10}
{Su}, M., {Slatyer}, T.~R., \& {Finkbeiner}, D.~P. 2010, \apj, 724, 1044

\bibitem[{{Veilleux} {et~al.}(2005){Veilleux}, {Cecil}, \&
  {Bland-Hawthorn}}]{veilleux05}
{Veilleux}, S., {Cecil}, G., \& {Bland-Hawthorn}, J. 2005, \araa, 43, 769

\end{thebibliography}
